# High-dimensional Array Bayesian Screening Based on Distributions with Structural Zeroes

**A. Lawrence Gould[1*], Erina Paul[1], Piyali Basak[1], Arinjita Bhattacharyya[1], Himel Mallick[2,3*]**

[1]Biostatistics and Research Decision Sciences, Merck & Co., Inc., Rahway, NJ 07065, USA

[2]Division of Biostatistics, Department of Population Health Sciences, Weill Cornell Medicine, Cornell University, New York, NY 10065, USA

[3]Department of Statistics and Data Science, Cornell University, Ithaca, NY 14850, USA

*** Correspondence:**
him4004@med.cornell.edu; larry_gould@merck.com

## Abstract

In many biomedical applications with high-dimensional features, such as single-cell RNA-sequencing, it is not uncommon to observe numerous structural zeros. Identifying important features from a pool of high-dimensional data for subsequent detailed analysis is often of interest. Here, we describe an exact, rapid Bayesian screening approach with attractive diagnostic properties, utilizing a Tweedie model. The method provides the likelihood that a feature with structural zeros merits further investigation, as well as distributions of the effect magnitudes and the proportion of features with the same expected responses under alternative conditions. The method is agnostic to assay, data type, and application. Through numerical studies, we demonstrate that the proposed methodology is effective in identifying important features for follow-up experimentation across a range of applications, including single-cell differential expression analysis of embryonic stem cells and embryonic fibroblasts in mice and differential analysis of CD4 and CD8 Peripheral Blood Mononuclear Cells (PBMCs) in humans.

## 1 Introduction

Typical omics assays generate matrices where rows correspond to features in a high-throughput array and columns represent omics measurements taken under ‘control’ and ‘test’ conditions. [1] Zero values are common and can arise due to technical factors (e.g., detection limits, sequencing depth) or biological variation (true absence in a sample). For example, in single-cell RNA sequencing, excess zeros are particularly challenging due to dropout effects, where lowly expressed genes may not be detected, leading to an overrepresentation of zero values. [1] A common method to address these zeros, particularly before log transformation or in compositional data analysis, is by adding a small value, known as a pseudo-count. [1] While this approach may seem reasonable and straightforward, it is ultimately ad hoc. Alternative methods involve modeling excess zeros using probability models, such as zero-inflated count regression, which explicitly accounts for two types of zeros: structural (biological) and sampling (technical). [2] However, in practice, distinguishing these sources of zeros can be challenging, and incorrectly specifying the zero-generating mechanism may lead to misinterpretations. [3] Recently, Tweedie models, which can accommodate both continuous and discrete components while handling excess zeros, have been applied to various omics assays, including single-cell and microbiome sequencing counts. [1]

The Tweedie distribution offers a particularly flexible approach to handling data with excess zeros, as it can model both a discrete probability mass at zero and a continuous distribution over positive values. [1] This makes it especially suitable for scenarios where zeros arise from different sources, such as structural zeros—occurring due to inherent biological absence—and sampling zeros, which result from detection limits or other technical factors. Unlike many other distributions that require separate components or adjustments to account for these zeros, the Tweedie distribution accommodates them within a unified framework, enabling more accurate and interpretable modeling of complex omics data with varying zero patterns.

We propose in this paper a novel screening procedure for high-dimensional omics arrays with excess zeros. The goal is to reduce a large collection of omics features to a manageable number of potentially “interesting” features while maintaining an acceptably small probability of missing any “interesting” ones. Mallick et al. [4] recently reviewed a wide range of statistical methods and software for analyzing microbiome data, which are characterized by zero inflation, overdispersion, high dimensionality, and sample heterogeneity. Methods for Bayesian screening calculations have been previously described when the ‘control’ and ‘test’ observations for any row of the array are scalars, and their corresponding likelihoods are Gaussian, binomial, or Poisson. **[5-7]** We extend Bayesian screening calculations to cases where the observations are not scalars or when the likelihoods are neither Gaussian, binomial, nor Poisson, or even expressible in closed form. The calculations are illustrated in the case where the observations are drawn from Tweedie distributions. **[8-11]** Marginal likelihoods for the 'test' observations are obtained conditionally on the corresponding 'control' observations, assuming as a prior the posterior distribution of the 'control' observations, either directly or with a defined shift. The resulting posterior probabilities determine whether the 'test' observations are similar to the 'control' observations. We demonstrate these calculations using observations drawn from Tweedie distributions. The proposed method is agnostic to assay, data type, and application.

While conceptually similar to traditional differential analysis approaches, our method has several key differences. Instead of focusing on detecting features that differ between experimental conditions while controlling the false discovery rate (FDR), as is common in most methods, our approach provides a statistically valid tool for subject matter experts to identify and, if appropriate, remove features unlikely to differ materially between experimental conditions. This shifts the key metric to the missed discovery rate (MDR). **[5, 6]** Reducing the dimensionality of the problem enhances the efficiency and effectiveness of methods specifically designed to identify important features.

When experimental conditions can be formally identified as "control" or "test," the posterior distributions of metrics expressing differences between the conditions can be conditioned on the 'control' feature response. The key issue is whether the process generating the expected response of a feature to the "test" condition is the same as—or differs from—the process generating the expected feature response to the 'control' condition. Our approach to determining prior distributions of key model parameters is novel, as it explicitly incorporates clinical and regulatory requirements regarding acceptable error rates. The calculations are exact (non-asymptotic) and rapid, as they do not require Markov Chain Monte Carlo (MCMC) computations.

The paper is structured as follows: We first introduce the method and define the underlying generative model. We then describe the Bayesian screening calculations for Tweedie distributed omics measurements and provide two real-life examples using single-cell data. Finally, we discuss the benefits and limitations of the method, as well as opportunities for future research. An annotated example of the R code usage and a full listing of the R code are provided in the Supplemental Information.

## 2 Method

### A *Data Structure*

A typical omics experiment yields a matrix of N outcomes where (for example) the rows correspond to the omics features (e.g., genes) in N samples (e.g., cells) of a screening array (e.g., single-cell RNA-sequencing) and the columns correspond to 1 or more measurements (observations) made under control and test regimes,

$$\mathbf{X}_C = \begin{bmatrix} \vec{x}'_{C1} \\ \vdots \\ \vec{x}'_{CN} \end{bmatrix} \text{ and } \mathbf{X}_T = \begin{bmatrix} \vec{x}'_{T1} \\ \vdots \\ \vec{x}'_{TN} \end{bmatrix}.$$

'Control' and 'test' are only formal labels for distinguishing the sets of observations. The elements of $\mathbf{X}_C$ and $\mathbf{X}_T$ usually will be non-negative real numbers. Without loss of generality, we will refer to omics features as genes, although the approach is broadly applicable beyond genes or genomic features. Similarly, we refer to cells as samples, which could also represent spots in spatial omics, biospecimens in microbiome studies, and other omics assays.

An observation vector for gene i is a realization from a joint distribution $f(\vec{x}_{Ci}; \vec{\theta}_{Ci})$ or $f(\vec{x}_{Ti}; \vec{\theta}_{Ti})$. The gene-specific parameters $\vec{\theta}_{Ci}$ of the joint distribution of the 'control' observations are realizations from a population process $g(\vec{\theta}_{Ci}; \vec{\Theta}_C)$ that is not gene-specific. The gene-specific parameters $\vec{\theta}_{Ti}$ of the joint distribution of the 'test' observations might be realizations from the 'control' population process $g(\vec{\theta}_{Ti}; \vec{\Theta}_C)$, or from a 'test' population process $g(\vec{\theta}_{Ti}; \vec{\Theta}_T)$ where $\vec{\Theta}_T \neq \vec{\Theta}_C$. The aim of the screening process is to determine for each gene whether the gene-specific parameter $\vec{\theta}_{Ti}$ is a realization from $g(\vec{\theta}_{Ti}; \vec{\Theta}_C)$ or $g(\vec{\theta}_{Ti}; \vec{\Theta}_T)$.

When $\vec{x}_{Ci}$ and $\vec{x}_{Ti}$ are scalars ($x_{Ci}$, $x_{Ti}$, i = 1,…,N), the likelihoods for $x_{Ci}$, and $x_{Ti}$ are assumed to have the same functional form $f(x; \vec{\theta}_i)$ with $\vec{\theta}_i = \vec{\theta}_{Ci}$ or $\vec{\theta}_{Ti}$. When $\vec{x}_{Ci}$ and $\vec{x}_{Ti}$ are not scalars, the elements of an observation vector $\vec{x}_i$ ($\vec{x}_{Ci}$ or $\vec{x}_{Ti}$) may or may not be independently distributed. If they are not independently distributed, then their joint distribution could be multivariate (e.g., multivariate normal), or could take the form of a copula incorporating the marginal densities of the individual elements. If they are independent, the likelihoods for $\vec{x}_{Ci}$ and $\vec{x}_{Ti}$,

$$f(\vec{x}_{Ci};\vec{\theta}_{Ci}) = \prod_k f_k(x_{Cik};\vec{\theta}_{Cik}), \quad f(\vec{x}_{Ti};\vec{\theta}_{Ti}) = \prod_k f_k(x_{Tik};\vec{\theta}_{Tik})$$

could have different functional forms from one element to the next ($f_k \not\equiv f$), or the same functional form with possibly different parameter values ($f_k \equiv f$). In this latter case, the 'control' and 'test' observation vectors could have different numbers of elements.

This paper describes the Bayesian screening process in general terms. It addresses in detail the case when the elements of the observation vectors have independent Tweedie distributions and possibly different parameter values ($f_k \equiv f$). 'Control' and 'test' refer to the effect of different regimes on the expression of samples from a particular type of cell. For example, this could include samples of lymphocytes from patients with different leukemia etiologies, samples of specific cells responding to different drugs, or independent and identically distributed (iid) replicates of observations on different cell types, such as stem cells versus fibroblasts.

*B Strategy Overview*

In general, marginal likelihoods for the 'test' observations are obtained for each row of the data array (each 'gene'), conditional on the corresponding 'control' observations. This is done under one of two assumptions: either the posterior distribution of the parameters for the 'control' observations serves as the prior distribution for the parameters of the 'test' observation distributions, or the prior distribution of the 'test' observation distributions is shifted in a defined way from the posterior distribution of the parameters of the 'control' observation distributions. The posterior probability associated with either of these possibilities, based on the 'test' observations, forms the basis for assessing whether the 'test' observations are or are not 'like' the 'control' observations.

Appendix 1 describes the Bayesian screening process in general terms, applicable whether the observations in the 'control' or 'test' regime are scalar or vector quantities, and whether they are independent or correlated. Appendix 2 addresses computational considerations when the likelihoods of the observations do not have conjugate priors or even closed-form expressions. Appendix 3 describes metrics for assessing differences between scalar and vector observation distributions when the ‘control’ and ‘test’ regime likelihoods are generated by the same process and when they are not, particularly when they are non-Gaussian. Appendix 4 provides a log of an actual software implementation, including results.

*C   Tweedie Distributions*

A number of approaches for analyzing zero-inflated non-negative continuous data have been considered in the literature. **[4, 12-14]** Tweedie distributions **[15]** provide a convenient way to accommodate the possibility that the distribution of experimental outcomes has positive mass at zero, which occurs for the dataset used here to illustrate the calculations (**Fig. 1**). Tweedie distributions are generalized exponential distributions that are characterized by a special mean-variance relationship:

$$\text{Variance} = \text{Dispersion} \times \text{Mean}^{\text{Power}} \quad (V = \phi\alpha^{\xi}).$$

Closed form expressions for the density of a Tweedie distribution exist only for special cases. Specific values of $\xi$ correspond to standard distributions (0 $\leftrightarrow$ Normal, 1 $\leftrightarrow$ Poisson, 2 $\leftrightarrow$ Gamma). The distributions are not defined for $0 < \xi < 1$. Values of $\xi$ between 1 and 2 correspond to compound Poisson-Gamma distributions that can have positive mass at zero. The parameters for any observation are the elements of a three-element vector, $\omega = (\xi, \alpha, \rho)$. It is more convenient computationally to work with transformations of these parameters that map to a vector $\mu$ in unbounded 3-space. The transformed values are $\mu_1 = \text{logit}(\xi - 1)$, $\mu_2 = \log(\alpha)$, $\mu_3 = \log(\phi)$ so that $\xi = 1 + (1 + e^{-\mu_1})^{-1}$, $\alpha = e^{\mu_2}$, and $\phi = e^{\mu_3}$.

When the parameters are estimated by maximum likelihood, the prior distribution of the parameters $\mu$ is asymptotically trivariate normal with mean $\theta$ and covariance matrix $\mathbf{V}_C$ under mild regularity conditions. The integral of the product of the prior distribution and the density of a control observation can be calculated using Gauss-Hermite quadrature with ML estimates $\hat{\theta}_C$ and $\hat{V}_C$ to give the marginal density of the control observations. The quadrature employed here uses a grid of points in 3-space with corresponding weights, obtained via a multivariate extension of the *gauss.hermite()* function in the *ecoreg* R package. **[16, 17]** A slight variation also produces the marginal density of a test observation when the test and control observations are generated by the same process, and the posterior mean is $\theta_{Ci}^{(0)}$. This is done separately for each gene.

When the same process does not generate the outcomes for the test and control features, values of the parameters for the 'control' group ($\xi$, $\alpha$, and $\phi$) are replaced by ($\xi^*$, $\alpha^*$, and $\phi^*$), where $\alpha^* = \mu_2 + \delta$, $\phi^* = \rho\phi$, and $\xi^*$ is defined by the odds ratio

$$\psi = \frac{(\xi^* - 1)(2 - \xi)}{(\xi - 1)(2 - \xi^*)}.$$

New transformed values ($\mu^*$) are calculated using ($\xi^*$, $\alpha^*$, $\phi^*$) in place of ($\xi$, $\alpha$, $\phi$) for the multivariate mean. The 'prior' distribution for calculating posterior probabilities corresponding to the 'test' observations for any gene is taken as trivariate normal with mean $\mu^*$ and covariance matrix $\widehat{V}_C$. New weights are calculated using the *gauss.hermite()* function and posterior marginal densities are calculated as before for the 'test' observations when the control process does not generate the test observations.

*D Metrics*

The computations described in detail in Appendix 1 provide the posterior probability that the same process generated the parameters for the likelihoods of the 'test' and 'control' observations of any gene. Low values of this probability identify genes that may be affected differently under the 'control' and 'test' regimes. A low value of the posterior probability with only a modest test-control difference can occur, so that perspective on the importance of potential differences needs additional consideration of the magnitudes of the differences.

If the observations are scalars ($\vec{x}_{Ci} = x_{Ci}$ and $\vec{x}_{Ti} = x_{Ti}$), then the difference between the 'control' and 'test' observation values for gene i (i = 1, …, N) can be expressed in various ways, of which the arithmetic difference $D_i = X_{Ti} - X_{Ci}$ or the ratio $R_i = X_{Ti}/X_{Ci}$ (assuming $X_{Ci} > 0$) ordinarily will be most useful. The complement of the cdf of either metric ("survival function") can provide useful insight, in particular quantifying the posterior probability that a difference or ratio exceeds one or more target values. Since Tweedie distributions can have positive probability at zero, it may be relevant to consider the probability that a ‘test’ outcome exceeds a target value conditional on a zero ‘control’ outcome, or conditional on a nonzero ‘control’ outcome, or unconditionally.

Useful definitions for the difference metric form are less obvious when there are vectors of ‘control’ and ‘test’ observations for each gene. Variations based on the Minkowski norm,

$$L_r(\vec{x}_{Ti}, \vec{x}_{Ci}) = \left(\vec{\mathbf{1}}' | \vec{x}_{Ti} - \vec{x}_{Ci}|^r\right)^{1/r}$$

are commonly used, especially the $L_2$ or Euclidean norm. **[18, 19]** The distribution of the metric (regardless of r) depends on the distribution of the observations. The $L_2$ norm has a chi-square distribution only when the observations are normally distributed, so is unlikely to be useful here because the observations do not follow gaussian distributions.

For simplicity of exposition and interpretation it is useful to assume that the components of any observation vector have the same likelihood, i.e., comprise sets of realizations from a common distribution that will depend on the gene and on the regime (‘control’ or ‘test’). This assumption implies that the observations for a gene are realizations from the same Tweedie distribution (separately for the ‘control’ and ‘test’ regimes). The assumption also reduces the comparison between vectors to a comparison between scalars, so that the approach used for scalar observations also can be used for vector outcomes.

*E Data*

The calculations are illustrated using observations from two datasets, described in **[1].** The first dataset (Islam) consists of a subset of gene expression measurements from replicates of single mouse embryonic stem cells and embryonic fibroblasts. The original dataset is available from the Gene Expression Omnibus (GEO) under accession GSE29087. The complete data file is an array with 22,936 rows (genes) and 92 columns (48 stem cells, 44 fibroblasts). The analyses presented here were conducted using a 1% random sample of the genes (229) and 20% random samples of each of the two cell types (9 fibroblasts, 10 stem cells). **Fig. 1** displays histogram plots of the values for typical fibroblasts and stem cells.

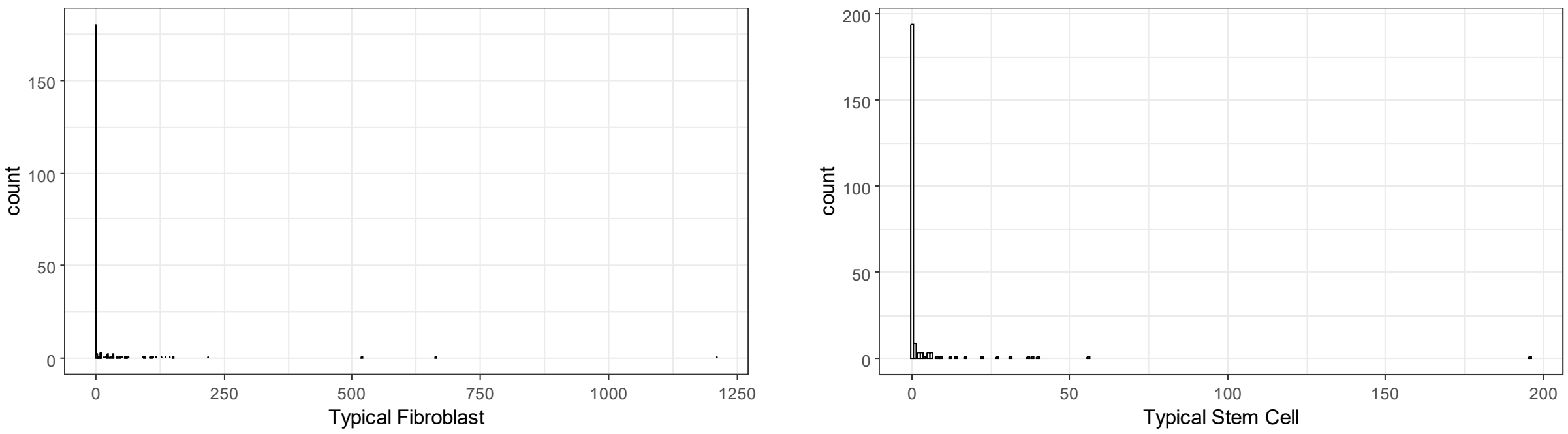

**Figure 1. Distributions of values of typical fibroblasts and stem cells in example Islam dataset.**

The second dataset (PBMC) consists of 13,713 genes and 2,638 cells collected from peripheral blood mononuclear cells (PBMCs). The goal is to distinguish CD4 cells from CD8 cells, which is a difficult problem due to their similar gene expression profiles, the complexity of immune cell differentiation, and potential technical noise in single-cell measurements. **[1]**.

## 3 Results

Since the 'test' observations on any gene could be generated by the same process that generated the 'control' observations, or a different process, an important quantity for gene-wise screening is the posterior probability $\pi_0$ that the same process applied for the observations on that gene. **Fig. 2** displays the posterior densities and cdfs of $\pi_0$ when fibroblasts are assumed to be the 'control' regime and stem cells are assumed to be the 'test' regime, and vice versa.

More importantly, the screening process identifies genes that may be expressed differently in the two cell types, as indicated in **Table 1**. The process is designed to detect differences reflecting increased expression levels, so that either regime could be a 'control'. **Table 1** illustrates this possibility because, at least for some genes (in particular, St3gal2 and 2010107H07Rik), the expressions for fibroblasts are substantially different from the expressions for stem cells, leading to very low probabilities ($P_{SP}$) that the same process generated the parameters for the observation likelihoods under the 'control' and 'test' regimes. This finding is due to the occurrence of very large gene expression levels

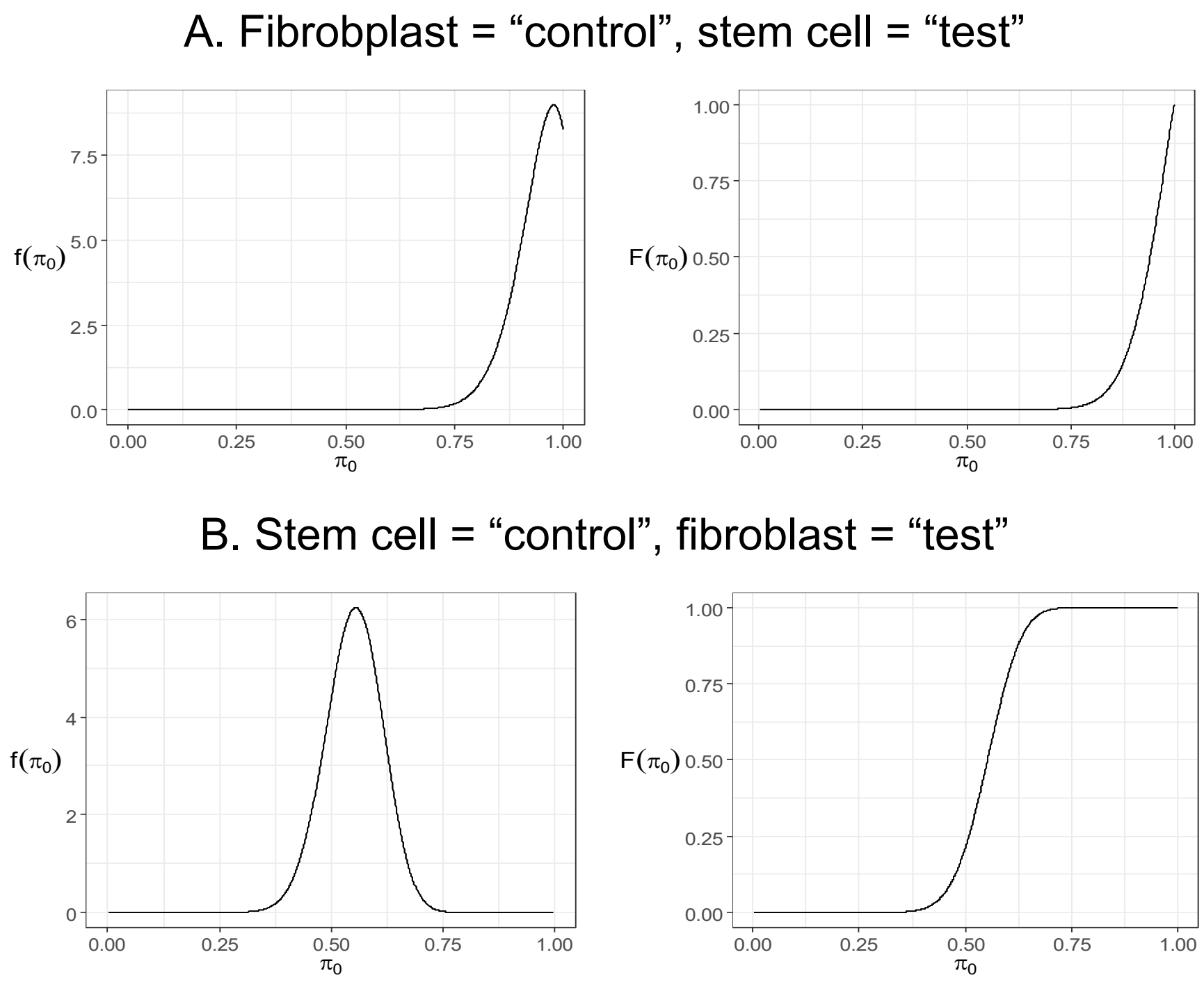

**Figure 2. Posterior density and cdf of $\pi_0$ depending on which cell type is ‘control’ for the Islam dataset (GSE29087).**

More importantly, the screening process identifies genes that may be expressed differently in the two cell types, as indicated in **Table 1**. The process is designed to detect differences reflecting increased expression levels, so that either regime could be a ‘control’. **Table 1** illustrates this possibility because, at least for some genes (in particular, St3gal2 and 2010107H07Rik), the expressions for fibroblasts are substantially different from the expressions for stem cells, leading to very low probabilities ($P_{SP}$) that the same process generated the parameters for the observation likelihoods under the ‘control’ and ‘test’ regimes. This finding is due to the occurrence of very large gene expression levels for at least one of the fibroblasts in the sample of fibroblasts included in the sample. On the other hand, it does not appear that gene expressions for stem cells were materially elevated relative to their expressions in fibroblasts. This is consistent with **Fig. 2** because the overall probability that the same process applied for the ‘control’ and ‘test’ regimes is much less when stem cells are the control than when fibroblasts are.

**Table 1. Posterior probabilities ($P_{SP}$) that the parameters for 'control' and 'test' regime observation likelihoods are generated by the same process in the Islam dataset. LR = Likelihood Ratio.**

| | Control = Fibroblasts | | Control = Stem Cells | |
|---|---|---|---|---|
| Gene | LR | $P_{SP}$ | LR | $P_{SP}$ |
| Snord2 | 0.738 | 0.947 | 0.488 | 0.713 |
| 9330175E14Rik | 0.738 | 0.947 | 0.488 | 0.713 |
| Dytn | 0.738 | 0.947 | 0.488 | 0.713 |
| St3gal2 | 0.967 | 0.932 | *19.98* | *0.06* |
| Itga8 | 0.885 | 0.938 | 0.897 | 0.577 |
| 2010107H07Rik | 1.308 | 0.912 | *488.0* | *0.003* |
| Napb | 0.739 | 0.947 | 0.523 | 0.699 |
| Mybl2 | 0.609 | 0.956 | 1.397 | 0.469 |
| Flt4 | 3.126 | 0.821 | 0.873 | 0.584 |

The difference metrics, as displayed in **Table 2** which assumes the stem cells are 'controls' and the fibroblasts are 'test', also are of interest. The table with the roles reversed is almost identical, and omitted. For the genes with a zero 'control' value, the probability that the corresponding 'test' value is much different from zero is quite small regardless of whether the 'control' and 'test' likelihood parameter values were generated by the same process or not. For the genes with a positive 'control' value, the probability of a materially larger 'test' value is appreciable (> 0.85), although in fact genes with a positive 'control' value were relatively scarce. Material differences regardless of the 'control' outcome also were relatively unlikely, although the probability was greater when the likelihood parameters were generated by different processes than by the same process.

**Table 2. Probabilities that the difference between an observation from a Tweedie distribution with "test" parameters generated by same process as "control" parameters and with "test" parameters generated by a different process in the Islam dataset.**

Stem Cells = "Control", Fibroblast = "Test"

| | Same Process for Control & Test | | | | Different Processes for Control & Test | | | |
|---|---|---|---|---|---|---|---|---|
| | P(Diff > d \| "Control" = 0) | | | | P(Diff > d \| "Control" = 0) | | | |
| Gene | d=20 | d=40 | d=60 | d=80 | d=20 | d=40 | d=60 | d=80 |
| Snord2 | 0.058 | 0.031 | 0.017 | 0.010 | 0.070 | 0.044 | 0.030 | 0.021 |
| 9330175E14Rik | 0.058 | 0.031 | 0.017 | 0.010 | 0.070 | 0.044 | 0.030 | 0.021 |
| Dytn | 0.058 | 0.031 | 0.017 | 0.010 | 0.070 | 0.044 | 0.030 | 0.021 |
| St3gal2 | 0.058 | 0.031 | 0.017 | 0.010 | 0.070 | 0.044 | 0.030 | 0.021 |
| Itga8 | 0.058 | 0.031 | 0.017 | 0.010 | 0.070 | 0.044 | 0.030 | 0.021 |
| 2010107H07Rik | 0.059 | 0.031 | 0.017 | 0.010 | 0.071 | 0.045 | 0.030 | 0.021 |
| Napb | 0.058 | 0.031 | 0.017 | 0.010 | 0.070 | 0.044 | 0.030 | 0.021 |
| Mybl2 | 0.062 | 0.035 | 0.020 | 0.012 | 0.072 | 0.046 | 0.032 | 0.023 |
| Flt4 | 0.058 | 0.031 | 0.017 | 0.010 | 0.070 | 0.044 | 0.030 | 0.021 |
| | P(Diff > d \| "Control" > 0) | | | | P(Diff > d \| "Control" > 0) | | | |
| Gene | d=20 | d=40 | d=60 | d=80 | d=20 | d=40 | d=60 | d=80 |
| Snord2 | 0.857 | 0.855 | 0.853 | 0.853 | 0.859 | 0.856 | 0.855 | 0.854 |
| 9330175E14Rik | 0.857 | 0.855 | 0.853 | 0.853 | 0.859 | 0.856 | 0.855 | 0.854 |
| Dytn | 0.857 | 0.855 | 0.853 | 0.853 | 0.859 | 0.856 | 0.855 | 0.854 |
| St3gal2 | 0.856 | 0.853 | 0.852 | 0.851 | 0.858 | 0.855 | 0.854 | 0.853 |
| Itga8 | 0.856 | 0.854 | 0.853 | 0.852 | 0.858 | 0.856 | 0.854 | 0.854 |
| 2010107H07Rik | 0.853 | 0.851 | 0.849 | 0.849 | 0.855 | 0.853 | 0.851 | 0.850 |
| Napb | 0.857 | 0.855 | 0.853 | 0.853 | 0.859 | 0.856 | 0.855 | 0.854 |
| Mybl2 | 0.854 | 0.852 | 0.851 | 0.850 | 0.856 | 0.854 | 0.852 | 0.851 |
| Flt4 | 0.856 | 0.853 | 0.852 | 0.851 | 0.858 | 0.855 | 0.854 | 0.853 |

**Table 2, continued**

| Gene | Stem Cells = "Control", Fibroblast = "Test" | | | | | | | |
|---|---|---|---|---|---|---|---|---|
| | Same Process for Control & Test | | | | Different Processes for Control & Test | | | |
| | P(Diff > d) | | | | P(Diff > d) | | | |
| | d=20 | d=40 | d=60 | d=80 | d=20 | d=40 | d=60 | d=80 |
| Snord2 | 0.055 | 0.029 | 0.016 | 0.009 | 0.067 | 0.042 | 0.029 | 0.020 |
| 9330175E14Rik | 0.055 | 0.029 | 0.016 | 0.009 | 0.067 | 0.042 | 0.029 | 0.020 |
| Dytn | 0.055 | 0.029 | 0.016 | 0.009 | 0.067 | 0.042 | 0.029 | 0.020 |
| St3gal2 | 0.055 | 0.029 | 0.016 | 0.009 | 0.067 | 0.042 | 0.029 | 0.020 |
| Itga8 | 0.055 | 0.029 | 0.016 | 0.009 | 0.067 | 0.042 | 0.029 | 0.020 |
| 2010107H07Rik | 0.056 | 0.029 | 0.016 | 0.009 | 0.068 | 0.043 | 0.029 | 0.020 |
| Napb | 0.055 | 0.029 | 0.016 | 0.009 | 0.067 | 0.042 | 0.029 | 0.020 |
| Mybl2 | 0.058 | 0.033 | 0.019 | 0.012 | 0.068 | 0.044 | 0.031 | 0.022 |
| Flt4 | 0.055 | 0.029 | 0.016 | 0.009 | 0.067 | 0.042 | 0.029 | 0.020 |

In the PBMC data analysis, the screening process identifies genes that may exhibit different expression levels between CD4 and CD8 cells, as shown in **Table 3**. Noteworthy genes in this example are 9130404D08Rik and Ogdh, where the expression levels in CD4 cells are significantly different from those in CD8 cells. This substantial difference leads to very low posterior selection probabilities ($P_{SP}$), indicating that the parameters for the observation likelihoods were unlikely to have been generated by the same process under the ‘control’ and ‘test’ conditions. The differences in the metrics presented in **Table 4**.

**Table 3. Posterior probabilities ($P_{SP}$) that the parameters for ‘control’ and ‘test’ regime observation likelihoods are generated by the same process in the PBMC data analysis. LR = Likehood Ratio.**

| Gene | Control = CD4 | | Control = CD8 | |
|---|---|---|---|---|
| | LR | $P_{SP}$ | LR | $P_{SP}$ |
| Tubb3 | 2.637 | 0.297 | 0.193 | 0.99 |
| 9130404D08Rik | 6.273 | 0.152 | 5.46 | 0.786 |
| Gm885 | 0.558 | 0.659 | 0.897 | 0.955 |
| Arg1 | 0.887 | 0.551 | 0.942 | 0.953 |
| Cenpv | 0.558 | 0.659 | 0.897 | 0.955 |
| Rtel1 | 0.602 | 0.642 | 0.624 | 0.968 |
| Fam169a | 0.558 | 0.659 | 0.897 | 0.955 |
| Abra | 0.558 | 0.659 | 0.897 | 0.955 |
| Exoc4 | 1.67 | 0.398 | 0.113 | 0.994 |
| Ogdh | 14.96 | 0.07 | 0.295 | 0.984 |

**Table 4. Probabilities that the difference between an observation from a Tweedie distribution with "test" parameters generated by same process as "control" parameters and with "test" parameters generated by a different process in PBMC data.**

| | CD4 = "Control", CD8 = "Test" | | | | | | | |
|---|---|---|---|---|---|---|---|---|
| | Same Process for Control & Test | | | | Different Processes for Control & Test | | | |
| | P(Diff > d \| "Control" = 0) | | | | P(Diff > d \| "Control" = 0) | | | |
| Gene | d=20 | d=40 | d=60 | d=80 | d=20 | d=40 | d=60 | d=80 |
| Tubb3 | 0.056 | 0.036 | 0.024 | 0.016 | 0.222 | 0.186 | 0.16 | 0.139 |
| 9130404D08Rik | 0.057 | 0.036 | 0.024 | 0.016 | 0.221 | 0.185 | 0.159 | 0.138 |
| Gm885 | 0.056 | 0.036 | 0.024 | 0.016 | 0.22 | 0.184 | 0.158 | 0.138 |
| Arg1 | 0.056 | 0.036 | 0.024 | 0.016 | 0.221 | 0.185 | 0.159 | 0.138 |
| Cenpv | 0.057 | 0.036 | 0.024 | 0.016 | 0.22 | 0.184 | 0.158 | 0.138 |
| Rtel1 | 0.058 | 0.037 | 0.026 | 0.018 | 0.221 | 0.185 | 0.159 | 0.138 |
| Fam169a | 0.056 | 0.036 | 0.024 | 0.016 | 0.22 | 0.184 | 0.158 | 0.138 |
| Abra | 0.056 | 0.036 | 0.024 | 0.016 | 0.22 | 0.184 | 0.158 | 0.138 |
| Exoc4 | 0.057 | 0.036 | 0.024 | 0.016 | 0.223 | 0.187 | 0.16 | 0.139 |
| Ogdh | 0.057 | 0.036 | 0.024 | 0.016 | 0.223 | 0.187 | 0.16 | 0.139 |

## 4. Discussion

This paper describes a process for screening entities, such as genes, to identify those whose expressions differ when the expressions consist of groups of observations that may reflect different interventions (treatments) or different entities, such as cell types (as in the Islam and the PBMC datasets), referred to as 'regimes.' The observations are assumed to be realizations from Tweedie distributions to account for the possibility that many expression values may be zero. The parameter values and the number of observations under each regime may vary. The parameters for the individual observation likelihoods are assumed to be exchangeable, i.e., realizations from a common regime-specific distribution. These assumptions simplify interpretation, allow for the quantification of observational differences, and limit the number of parameters.

The formulation of the screening process depends only on Bayes factors calculated from predictive likelihoods, meaning that most of these assumptions, while convenient, are not necessary. Thus, Tweedie distributions are useful but not essential. Additionally, suppose that there are an equal number of 'control' and 'test' observations, that the observations within each regime are correlated, and that their joint distributions are described by copulas with potentially different functional forms for the marginal distributions. Although this scenario could introduce many more parameters and significantly increase the computational burden, predictive likelihoods could still be calculated, enabling the screening process described in Appendix 1 to be carried out.

Beyond gene expression analysis, the proposed screening methodology may be applicable to other high-dimensional biological datasets, such as microbiome, proteomics, or metabolomics, where sparsity, dependence structures, and heterogeneous regimes play a critical role in inference. Further exploration of alternative distributional assumptions and inference strategies may enhance the applicability of this approach in broader scientific settings.

R code for carrying out the calculations is provided in the Supplementary Information.

## CONFLICT OF INTEREST

The authors report no conflict of interest.

## **APPENDIX 1** Bayesian Screening

Bayesian screening methods based on a random mixing parameter that selects one of two potential likelihoods for observed outcomes have been studied previously (e.g., **[6])**. The random mixing parameter that these methods use provides an automatic adjustment to reduce the potential for 'false alarms'**.[20, 21]** The following discussion formalizes and generalizes the development for normally distributed outcomes. **[6]**

The following lemma provides expressions for key functionals of $\mathbf{X}_C$ and $\mu_C$. when the control regime values $\vec{x}_{Ci}$, i = 1, …, N, are exchangeable and the corresponding parameters $\mu_{Ci}$ can be regarded as realizations from a prior density $f_\mu\left(\mu_{Ci};\ \theta_C^{(0)}\right)$ with common (vector) parameter $\theta_C^{(0)}$. This is a testable assumption. **[22]**

<u>LEMMA 1</u>. Let $\mathbf{X}_C$ denote a matrix whose N rows $\vec{x}'_{C1}$, i = 1, …, N comprise measurements ('features') made under a 'control' regime. Let $f_X(\vec{x}_{Ci};\ \vec{\mu}_{Ci})$ denote the likelihood corresponding to $\vec{x}_{Ci}$ and let $\mu_C$ denote a matrix whose rows are $\vec{\mu}'_{Ci}$, i = 1, …, N. If the rows of $\mathbf{X}_C$ are independently distributed and exchangeable, then

(a) the marginal likelihood for $\vec{\mathbf{x}}_{Ci}$ is

$$g_X\left(\vec{x}_{Ci};\theta_C^{(0)}\right) = \int f_X(\vec{x}_{Ci};\vec{\mu})f_\mu\left(\vec{\mu};\theta_C^{(0)}\right)d\vec{\mu} \tag{1a}$$

(b) the posterior density of $\vec{\mu}_{Ci}$ is

$$h_\mu\left(\vec{\mu}_{Ci};\vec{x}_{Ci},\theta_C^{(0)}\right) = f_X(\vec{x}_{Ci};\vec{\mu}_{Ci})f_\mu\left(\vec{\mu}_{Ci};\theta_C^{(0)}\right)/g_X\left(\vec{x}_{Ci};\theta_C^{(0)}\right) \tag{1b}$$

(c) the posterior expected value of $\mu_{Ci}$ is

$$\tilde{\theta}_{Ci} = \int \vec{\mu}\, f_X(\vec{x}_{Ci};\vec{\mu})f_\mu\left(\mu;\theta_C^{(0)}\right)d\vec{\mu}/g_X\left(\vec{x}_{Ci};\theta_C^{(0)}\right) \tag{1c}$$

(d) the predictive likelihood of a future value of $\vec{x}_{Ci}$ given an observed value $\vec{x}_{Ci}^{(obs)}$ and $\theta_C^{(0)}$ is

$$\tilde{g}_X(\vec{x}_{Ci};\tilde{\theta}_{Ci}) = \int f_X\left(\vec{x}_{Ci}^{(new)};\ \vec{\mu}\right)h_\mu\left(\vec{\mu};\vec{x}_{Ci}^{(obs)},\theta_C^{(0)}\right)d\vec{\mu} \tag{1d}$$

<u>Proof</u>: By construction.

COROLLARY 1. If $\mathbf{X}_C$ denotes a matrix whose N rows $\vec{x}'_{T1}$, i = 1, …, N comprise measurements made under a 'test' regime and the control and test regimes have indistinguishable effects on the expression of the gene corresponding to the i-th row of the screening array, then (1d) also is the predictive likelihood of a future value of $\vec{x}_{Ti}$ given an observed value $\vec{x}_{Ci}^{(obs)}$ after replacing $\vec{x}_{Ci}^{(new)}$ with $\vec{x}_{Ti}^{(new)}$.

Proof: By construction.

The test regime values $\vec{x}_{Ti}$, i = 1, …, N, may not be exchangeable across the rows of the array because the regime may affect the different genes differently, so that the prior density of $\mu_{Ti}$ given the 'control' regime observations is $f_\mu\left(\mu_{Ti}; \theta_{Ti}^{(0)}\right)$ where $\theta_{Ti}^{(0)} \neq \tilde{\theta}_{Ci}$.

LEMMA 2. If the control and test regimes have different effects on the expression of the gene corresponding to $\vec{x}_{Ti}$ and $\tilde{\theta}_{Ti}^{(0)} = q(\tilde{\theta}_{Ci})$, where q() is a known function, then the predictive likelihood for $\vec{x}_{Ti}$ given an observed control regime outcome, $\vec{x}_{Ci}^{(obs)}$, is

$$g_X\left(\vec{x}_{Ti}; \tilde{\theta}_{Ti}^{(0)}\right) = f_X(\vec{x}_{Ti};\ \vec{\mu}_{Ti}) f_\mu\left(\vec{\mu}_{Ti};\ \tilde{\theta}_{Ti}^{(0)}\right) / \int f_X(\vec{x}_{Ti};\ \vec{\mu}) f_\mu\left(\vec{\mu};\ \tilde{\theta}_{Ti}^{(0)}\right) d\vec{\mu} \qquad (2)$$

Proof: By construction.

Remark: The use of $\tilde{\theta}_{Ci}$ instead of $\theta_C^{(0)}$ 'calibrates' the response to the test regime for gene i to the corresponding response to the control regime.

Whether the control and test regimes have the same or different effects on the expression of the i-th gene is unknown. Suppose that the actual parameter $\theta_{Ti}^{(0)}$ of the prior distribution of $\mu_{Ti}$, i = 1, …, N, is a random mixture of $\tilde{\theta}_{Ci}$ and $\tilde{\theta}_{Ti}^{(0)}$ with random mixing parameter $\gamma_i$,

$$\theta_{Ti}^{(0)} = (1 - \gamma_i)\tilde{\theta}_{Ci} + \gamma_i \tilde{\theta}_{Ti}^{(0)} \qquad (3)$$

so that $\gamma_i$ = 0 implies that the same process produced the control and test parameters and $\gamma_i$ = 1 implies that different processes produced the parameters and let $\gamma = (\gamma_1, \ldots, \gamma_N)$. We assume that each $\gamma_i$ is a realization from a Bernoulli distribution with parameter $\pi_0$ that expresses the *a priori* probability that the same process operates for the control and test observations for any gene. We incorporate $\pi_0$ as an explicit model parameter. Strategies for determining the value of $\pi_0$ without the assumption of a prior distribution have been described in the hypothesis-testing literature. **[23-29]**

THEOREM. Let $\gamma_i$ have a Bernoulli probability function with parameter $\pi_0$, $p(\gamma_i;\, \pi_0) = \pi_0^{1-\gamma_i}(1-\pi_0)^{\gamma_i}$, and denote the prior density of $\pi_0$ by $f_{\pi_0}^{(0)}(\pi_0;\, \alpha)$, e.g., a beta density with parameters $(\alpha,1)$ [2,3]). Further, let $B_i = g_X(\vec{x}_{Ti};\, \tilde{\theta}_{Ti}^{(0)})/g_X(\vec{x}_{Ti};\, \tilde{\theta}_{Ci})$. Then

(a) the posterior density of $\pi_0$ is

$$f_{\pi_0}^{(post)}(\pi_0;\, \mathbf{X}_C, \mathbf{X}_T, \boldsymbol{\alpha}) = H^{-1}\pi_0^N f_{\pi_0}^{(0)}(\pi_0;\, \boldsymbol{\alpha}) \prod_{i=0}^N \left\{1 + \frac{1-\pi_0}{\pi_0} B_i\right\} \tag{4}$$

where $$H = \int_{\pi_0} \pi_0^N f_{\pi_0}^{(0)}(\pi_0;\, \boldsymbol{\alpha}) \prod_{i=0}^N \left\{1 + \frac{1-\pi_0}{\pi_0} B_i\right\} d\pi_0$$

(b) the conditional (on $\pi_0$) posterior probability function of $\gamma_i$ is

$$\mathrm{ppost}(\gamma_i;\, B_i, \pi_0) = \left\{1 - \gamma_i + \frac{1-\pi_0}{\pi_0}\gamma_i B_i\right\}\left\{1 + \frac{1-\pi_0}{\pi_0} B_i\right\}^{-1} \tag{5}$$

(c) the unconditional posterior probability function of $\gamma_i$ is

$$\tilde{p}_{post}(\gamma_i; B_i, \alpha) = \int p_{post}(\gamma_i;\, B_i, \pi_0)\, f_{\pi_0}^{(post)}(\pi_0;\, \mathbf{X}_C,\, \mathbf{X}_T,\, \alpha) d\pi_0 \tag{6}$$

Proof:

The product of the likelihoods and prior densities for all of the genes is

$$\begin{aligned} f\left(\mathbf{X}_C, \vec{x}_{Ti}, \gamma, \pi_0; \alpha, \theta_C^{(0)}, \tilde{\theta}_{Ti}^{(0)}, \mu_C, \mu_T\right) &= f_{\pi_0}^{(0)}(\pi_0;\, \alpha) \prod_{i=1}^N f_X(\vec{x}_{Ci};\, \vec{\mu}_{Ci})\, f_\mu\left(\mu_{Ci};\, \theta_C^{(0)}\right) \\ &\times \prod_{i=1}^N f_X(\vec{x}_{Ti};\, \vec{\mu}_{Ti})\, f_\mu\left(\mu_{Ti};\, (1-\gamma_i)\tilde{\theta}_{Ci} + \gamma_i\tilde{\theta}_{Ti}^{(0)}\right) \pi_0^{1-\gamma_i}(1-\pi_0)^{\gamma_i} \end{aligned} \tag{7}$$

Integrating (7) separately with respect to the members of $\mu_C$ and $\mu_T$ yields the joint marginal densities of the control and test observations on each feature,

$$\begin{aligned} g\left(\mathbf{X}_C, \mathbf{X}_T, \gamma, \pi_0; \alpha, \theta_C^{(0)}, \tilde{\theta}_T^{(0)}\right) &= f_{\pi_0}^{(0)}(\pi_0;\, \alpha) \prod_{i=1}^N g_X(\vec{x}_{Ci};\, \theta_C^{(0)}) \\ &\times \prod_{i=1}^N \left\{\pi_0^{1-\gamma_i}(1-\gamma_i) g_X(\vec{x}_{Ti};\, \tilde{\theta}_{Ci}) + (1-\pi_0)^{\gamma_i}\gamma_i g_X(\vec{x}_{Ti};\, \tilde{\theta}_{Ti}^{(0)})\right\} \end{aligned} \tag{8}$$

Dividing (7) by (8) yields the conditional probability function of $\gamma_i$ (5). Summation of (8) separately with respect to each $\gamma_i$ yields the product of the joint marginal density of the observed values and the density of the probability ($\pi_0$) that any $\gamma_i$ is zero,

$$\begin{aligned} f\left(\mathbf{X}_C, \mathbf{X}_T, \pi_0; \alpha, \theta_C^{(0)}, \theta_T^{(0)}\right) = f_{\pi_0}^{(0)}(\pi_0;\ \boldsymbol{\alpha}) \prod_{i=1}^{N} g_X\left(\vec{x}_{Ci};\ \theta_C^{(0)}\right) \\ \times \prod_{i=1}^{N}\left\{\pi_0 g_X\left(\vec{x}_{Ti}; \tilde{\theta}_{Ci}\right) + (1-\pi_0) g_X\left(\vec{x}_{Ti}; \tilde{\theta}_{Ti}^{(0)}\right)\right\} \end{aligned} \qquad (9)$$

Integrating (9) with respect to $\pi_0$ gives the marginal density of the observations; dividing this quantity into (8) yields the posterior density of $\pi_0$ (4). The unconditional probability function of $\gamma_i$ (6) is the expectation of (5) with respect to (4). QED

## **APPENDIX 2**. Computational Considerations

The calculations are straightforward when the likelihoods have explicit algebraic forms and conjugate prior distributions. **[5, 6]** When the likelihoods do not have conjugate priors, or even closed expressions for the densities, the marginal densities of the observations and posterior expectations of the parameters must be determined numerically. This can be accomplished by framing the computations as standard Bayesian calculations and applying MCMC methods. A simpler and more rapid approach proceeds via Empirical Bayes and Gauss-Hermite integration to exploit the approximate normality of maximum likelihood estimators. **[16, 17, 30, 31]**

Suppose that the control regime likelihood values $f_X(\vec{x}_{Ci};\ \vec{\mu}_{Ci})$, i = 1,…,N can be calculated at least numerically. Convergence difficulties may arise with ML calculations when some of the components of $\vec{\mu}_{Ci}$ are bounded. Reparametrization avoids this problem by transforming the elements of $\vec{\mu}_{Ci}$ to another vector or scalar, say $\vec{\eta}_{Ci}$, whose elements are not bounded, so that no generality is lost in this application by assuming that the elements of elements of $\vec{\mu}_{Ci}$ are not bounded. In what follows, the values of $\vec{x}_{Ci}$ and the test regime observations $\vec{x}_{Ti}$ are assumed to be iid with respective parameter vectors $\vec{\mu}_{Ci}$ and $\vec{\mu}_{Ti}$. This assumption could be relaxed at least in principle, e.g., by the use of copulas.

If an acceptable prior distribution for $\vec{\mu}_{Ci}$ is available, then the calculations should use it. Otherwise, an Empirical Bayes approach that uses an asymptotic normal distribution of the maximum likelihood estimator (MLE) of $\vec{\mu}_{Ci}$ could be used as a prior distribution for $\mu_{Ci}$ as now is described. The MLE of $\mu_{Ci}$ is

$$\hat{\mu}_C = \underset{\vec{\mu}}{\text{argmax}} \left\{ \sum_{i=1}^{N} \log\left( f_X(\vec{x}_{Ci};\ \vec{\mu}) \right) \right\}$$

The corresponding asymptotic covariance matrix of $\hat{\mu}_C$ is $V_\mu$ and the prior density of $\vec{\mu}_{Ci}$ is multivariate normal with mean $\hat{\mu}_C$ and covariance matrix $V_\mu$. With this prior density, the marginal density of $\vec{x}_{Ci}$ can be calculated using Gauss-Hermite quadrature,

$$g_X\left(x_{Ci};\ \theta_C^{(0)}\right) = \int_{\vec{\mu}} f_{norm}(\vec{\mu}; \hat{\mu}_C, V_\mu) f_X(\vec{x}_{Ci};\ \vec{\mu}) d\vec{\mu} = \sum_{k=1}^{K} w_k\, f_X(x_{Ci};\ \mathbf{v}\vec{v}_k)$$

where $\vec{v}_1, \vec{v}_2, \dots, \vec{v}_K$ are points in $\mu$-space determined by the Gauss-Hermite quadrature procedure and $w_1, w_2, \dots, w_K$ are corresponding weights when the parameters of the prior density of $\boldsymbol{\mu}_{Ci}$ are $\hat{\mu}_C$ and $V_\mu$. The weights and quadrature points are obtained using functions in the ecoreg R package. **[16, 17]** The same calculation also yields the posterior mean $\tilde{\theta}_{Ci}$ for $\boldsymbol{\mu}_{Ci}$,

$$\tilde{\theta}_{Ci} = \sum_{k=1}^{K} w_k\, \vec{v}_k f_X(\vec{x}_{Ci};\ \vec{v}_k)$$

If the same process produces the values of the control and test group observations then, given the control group observations, the posterior mean $\tilde{\theta}_{Ci}$ can be taken as the parameter of the prior distribution of the test group $\boldsymbol{\mu}_{Ti}$ values. Consequently, the marginal density of the i-th feature value $\vec{x}_{Ti}$ when the same processes operate for the control and test interventions is obtained from

$$g_X(\vec{x}_{Ti};\ \tilde{\theta}_{Ci}) = \int_{\vec{\mu}} f_{norm}(\vec{\mu}; \tilde{\theta}_{Ci}, V_\mu) f_X(\vec{x}_{Ci};\ \vec{\mu}) d\vec{\mu} = \sum_{k=1}^{K} w_k\, f_X(\vec{x}_{Ti};\ \vec{v}_k) \qquad (12)$$

where $\vec{v}_1, \vec{v}_2, \dots, \vec{v}_K$ are points in $\mu$-space determined by the Gauss-Hermite quadrature procedure and $w_1, w_2, \dots, w_K$ are corresponding weights when the parameters of the prior density of $\vec{\mu}$ are $\tilde{\theta}_{Ci}$ and $V_\mu$.

Suppose that the expected value of $\vec{\mu}_{Ti}$ when the feature values for the interventions are not produced by the same process is a function $\tilde{\theta}_{Ti}^{(0)} = q(\tilde{\theta}_{Ci};\ \zeta)$ of the expectation when the values are produced by the same process. Suppose also that the prior density of $\vec{\mu}_{Ti}$ is a multivariate normal density with expectation $\tilde{\theta}_{Ti}^{(0)}$ and covariance matrix $V_\mu$. The marginal density of the i-th feature value $\vec{x}_{Ti}$ when different processes operate for the control and test interventions is obtained from

$$g_X\left(x_{Ti};\ \tilde{\theta}_{Ti}^{(0)}\right) = \sum_{k=1}^{K} w_k\, f_X(\vec{x}_{Ti};\ \vec{v}_k) \qquad (13)$$

where $\vec{v}_1, \vec{v}_2, \ldots, \vec{v}_K$ are points in μ-space determined by the Gauss-Hermite quadrature procedure and $w_1, w_2, \ldots, w_K$ are corresponding weights when the parameters of the prior density of $\mu_{Ti}$ are $\tilde{\theta}_{Ti}^{(0)}$ and $V_\mu$. The quantities needed to compute the cdf of the metric ($D_i$ or $R_i$) are computed using (12) and (13).

The posterior density (4) and cdf of $\pi_0$ can be calculated easily using (12) and (13) because the quantity H can be calculated by univariate numerical integration. The marginal posterior probability (6) that $\gamma_i = 0$ (i.e., that the same process produced the test and control feature values for gene i) is, following the Theorem in Appendix 1,

$$p_{post}(\gamma_i = 0; \mathbf{x}_C, \mathbf{x}_T, \alpha, \tilde{\theta}_{Ci}, \tilde{\theta}_{Ti}^{(0)}) = \int_{\pi_0} f_{\pi_0}^{(post)}(\pi_0;\ \mathbf{x}_C, \mathbf{x}_T, \alpha) \left\{1 + \frac{1-\pi_0}{\pi_0} B_i\right\}^{-1} d\pi_0$$

## **APPENDIX 3**. Test-Control Difference Metrics

Appendix 1 described the computation of the posterior probability that the same process generated the parameters for the likelihoods of the 'test' and 'control' observations. Low values of this probability identify genes that may be affected differently under the 'control' and 'test' regimes. However, it is possible to have a low value of the posterior probability with only a modest test-control difference, so that perspective on the importance of potential differences needs additional consideration of the magnitudes of the differences.

*Scalar observations*.

If the observations are scalars (so that $\vec{x}_{Ci} = x_{Ci}$ and $\vec{x}_{Ti} = x_{Ti}$), then the difference between the 'control' and 'test' observation values for gene i (i = 1, …, N) can be expressed in various ways, of which the arithmetic difference $D_i = X_{Ti} - X_{Ci}$ or the ratio $R_i = X_{Ti}/X_{Ci}$ (assuming $X_{Ci} > 0$) ordinarily will be most useful. The conditional cdf of the arithmetic difference between a test observation value $x_{Ti}$ and a corresponding control observation $x_{Ci}$ is

$$F_D\left(d; x_{Ci}, \tilde{\theta}_{Ti}^{(0)}\right) = \int_{-\infty}^{x_{Ci}+d} \tilde{g}_X\left(x; \tilde{\theta}_{Ti}^{(0)}\right) dx$$

(see Lemmas 1 or 2 in Appendix 1). The unconditional survival function (1-cdf) is

$$S_D\left(d; \tilde{\theta}_{Ti}^{(0)}, \theta_C^{(0)}\right) = P(D_i > d; \tilde{\theta}_{Ti}^{(0)}, \theta_C^{(0)}) = 1 - \int_x g_X(x;\ \theta_C^{(0)}) F_D\left(d; x, \tilde{\theta}_{Ti}^{(0)}\right) dx \qquad (10)$$

Similar expressions for the ratio are

$$F_R\left(r; x_{Ci}, \tilde{\theta}_{Ti}^{(0)}\right) = \int_{-\infty}^{r x_{Ci}} \tilde{g}_X\left(x; \tilde{\theta}_{Ti}^{(0)}\right) dx$$

$$S_R\left(r; \tilde{\theta}_{Ti}^{(0)}, \theta_C^{(0)}\right) = P(R_i > r;\ \tilde{\theta}_{Ti}^{(0)}, \theta_C^{(0)}) = 1 - \int_x g_X(x;\ \theta_C^{(0)}) F_R\left(r; x, \tilde{\theta}_{Ti}^{(0)}\right) dx \qquad (11)$$

The posterior distribution of M provides perspective on the magnitude of the difference for any gene. The conditional posterior cdf of M given $\gamma_i$ and the values of $\tilde{\theta}_{Ci}$ or $\tilde{\theta}_{Ti}^{(0)}$ (which depend on the observed feature values $x_{Ci}$ and $x_{Ti}$) is

$$P_{post}(M_i \le m;\ \gamma_i, \tilde{\theta}_{Ci}, \tilde{\theta}_{Ti}^{(0)}) = \left(1 - \gamma_i\right) F_M(m;\ \tilde{\theta}_{Ci}, \tilde{\theta}_{Ci}) + \gamma_i F_M(m;\ \tilde{\theta}_{Ci}, \tilde{\theta}_{Ti}^{(0)})$$

The unconditional posterior cdf of M is

$$P_{post}(M_i \leq m;\ \tilde{\theta}_{Ci}, \tilde{\theta}_{Ti}^{(0)}) = \tilde{p}_{post}\left(0;\ x_{Ci}, x_{Ti}, \alpha,\ \tilde{\theta}_{Ci}, \tilde{\theta}_{Ti}^{(0)}\right) \times F_M(m;\ \tilde{\theta}_{Ci}, \tilde{\theta}_{Ci}) \quad (12)$$

$$+ \tilde{p}_{post}\left(1;\ x_{Ci}, x_{Ti}, \alpha,\ \tilde{\theta}_{Ci}, \tilde{\theta}_{Ti}^{(0)}\right) \times F_M(m;\ \tilde{\theta}_{Ci}, \tilde{\theta}_{Ti}^{(0)})$$

*Vector observations*

Definitions for the form of the difference (or, more appropriate, distance) metric M are less obvious when there are vectors of 'control' and 'test' observations for each gene. Variations based on the Minkowski norm,

$$L_r(\vec{x}_{Ti}, \vec{x}_{Ci}) = \left(\vec{\mathbf{1}}' | \vec{x}_{Ti} - \vec{x}_{Ci} |^r\right)^{1/r}$$

are commonly used, especially the $L_2$ or Euclidean norm. **[18, 19]** The distribution of the metric (regardless of r) depends on the distribution of the observations. The $L_2$ norm has a chi-square distribution only when the observations are normally distributed.

Since the observations in the application considered here are not normally distributed, a different approach is needed. For reasons outlined in Appendix 3, which discusses computational considerations, we assume that the components of any observation vector have the same likelihood, i.e., comprise sets of realizations from a common distribution that will depend on the gene and on the regime ('control' or 'test'). The calculations described in Appendices 1 and 2 lead to values of the parameters of the likelihoods for each gene under each regime, from which the survival function values can be calculated using (10 – 12).

# R CODE FOR CALCULATIONS

```
GenScreenMenu.fn <- function(inits=c(1.5,2), Tmod=c(2,2,1), TargDiffs=c(20,40,60,80),
                             pi0X=".001*1:999", ngridpts=10, prune=0.2, digits=3, print.int=0)
{
  #######################################################################################
  #######################################################################################
  ###########################  Utility functions  #######################################
  #######################################################################################
  #######################################################################################

  string_to_num.fn <- function(ch_str)
  {
  #  This function converts a string representation of a numerical scalar or vector such
  #  as what the readline function provides to a numerical scalar or vector. For example,
  #  the strings  ch_str = "1 2 3 4" or "1, 2, 3, 4" or "1,2,3,4" produced by readline
  #  are converted to the numerical vector c(1,2,3,4).
  #                                                        A. L. Gould  February 2018
    xx <- stringr::str_replace_all(stringr::str_squish(ch_str),", ",",")
    return(eval(parse(text=paste0("c(",stringr::str_replace_all(xx," ",","),")"))))
  }

  getYorN.fn <- function(query)
      return(c("Y","N")[menu(c("Yes","No"),title=paste0("\n",query,"?"))])

  get_val.fn <- function(GVtext, default)
  {
    cat("\n")
    ifelse(length(default) > 1, xx <- paste0("c(",toString(default),")"), xx <- default )
    x <- readline(paste0("Enter ", GVtext, " (<CR> -> default value = ",xx, "): "))
    ifelse(!sjmisc::is_empty(x), return(string_to_num.fn(x)), return(default))
  }

  get_strval.fn <- function(GVtext, default=".001*1:999")
  {
    cat("\n")
    x <- readline(paste0("Enter ", GVtext, " (<CR> -> default value = ",default, "): "))
    ifelse(!sjmisc::is_empty(x), return(x), return(default))
  }
  ##################################################################################
  ##################################################################################
  #######                                                                 #########
  #######   Setup and execution of GenScreen calculations starts here     #########
  #######                                                                 #########
  ##################################################################################
  ##################################################################################

  cat("\n")
  rs <- readline("Enter a random seed integer value ")  # Get a random seed to
  eval(parse(text=paste0("set.seed(",rs,")")))          # insure reproducibility
  repeat
  {                                                    # Original dataset name
    cat("\n")
    dsn <- readline("What is name of the complete input dataset? ")
    if (!sjmisc::is_empty(dsn)) break
    else  print("Error: need to specify a dataset name")
  }
  origdta <- eval(parse(text=dsn))
  nrws <- nrow(origdta)
  ncls <- ncol(origdta)                                # Get indices of cntl & test obsns
  cat(paste0("\nEnter column indices (between 1 and ",ncls,") of ",dsn))
  x <- readline(paste0("corresponding to \'control\' obsns, separated by spaces or commas: "))
  ctl_cols <- string_to_num.fn(x)
  cat(paste0("\nEnter column indices (between 1 and ",ncls,") of ",dsn))
  x <- readline(paste0("corresponding to \'test\' obsns, separated by spaces or commas: "))
  tst_cols <- string_to_num.fn(x)
  repeat
  {                                                    # Specify fraction of rows of
    cat("\n")                                          # origdta to include in workdta
    x <- readline(paste0("Enter fraction of rows of ",dsn,
                         " to include in calculations: "))
```

```
    if (!sjmisc::is_empty(x)) break
    else  print("Error: need to specify a fraction value")
  }
  rfrac <- eval(parse(text=x))
  nrwrk <- round(rfrac*nrws)                          # No. of rows to include in workdta
  indx_rws <- sample(1:nrws, nrwrk)                   # Rows to include in workdta
  repeat
  {                                                   # Specify fraction of cntl cols of
    cat("\n")                                         # origdta to include in workdta
    x <- readline(paste0("Enter fraction of cntl columns of ",dsn,
                         " to include in calculations: "))
    if (!sjmisc::is_empty(x)) break
    else  print("Error: need to specify a fraction value")
  }
  ctlfrac <- eval(parse(text=x))
  nccwrk <- round(ctlfrac*length(ctl_cols))
  indx_ctl <- sample(ctl_cols, nccwrk)
  repeat
  {                                                   # Specify fraction of test cols of
    cat("\n")                                         # origdta to include in workdta
    x <- readline(paste0("Enter fraction of test columns of ",dsn,
                         " to include in calculations: "))
    if (!sjmisc::is_empty(x)) break
    else  print("Error: need to specify a fraction value")
  }
  tstfrac <- eval(parse(text=x))
  nctwrk <- round(tstfrac*length(tst_cols))
  indx_tst <- sample(tst_cols, nctwrk)
  xC <- origdta[indx_rws, indx_ctl]
  xT <- origdta[indx_rws, indx_tst]
  inits <- get_val.fn("initial values for ML estimation of prior prameters", inits)
  Tmod <- get_val.fn("elements of Tmod vector", Tmod)
  TargDiffs <- get_val.fn("target difference values", TargDiffs)
  pi0x <- get_strval.fn("points at which to compute density & cdf of pi0", pi0X)
  ngridpts <- get_val.fn("No. of gridpoints for gauss-hermite integration", ngridpts)
  prune <- get_val.fn("pruning constant for gauss-hermite integration", prune)
  digits <- get_val.fn("no. of digits for printing output", digits)
  print.notes <- get_val.fn("1 to print intermediate values by Engine,
                                          0 to suppress",print.int)==1
  run_engine <- getYorN.fn("Run screen engine now (just save specs list if no)")
  misclist <- list(rs=rs, nrws=nrws, ncls=ncls, rfrac=rfrac,indx_rws=indx_rws,
               ctlfrac=ctlfrac, indx_ctl=indx_ctl, tstfrac=tstfrac, indx_tst=indx_tst)
  GenScrnList <- list(xC=xC, xT=xT, inits=inits, Tmod=Tmod, TargDiffs=TargDiffs, pi0x=pi0x,
                         ngridpts=ngridpts, prune=prune, digits=digits, misc=misclist,
                         print.notes=print.notes)
  if(run_engine=="Y")  return(TweedieScreenEngine.fn(GenScrnList))
  else return(GenScrnList)
}

TweedieScreenEngine.fn <- function(inputlist)
{
# This function calculates a variety of quantities related to screening outcomes under
# 'control' (xC) and 'test' (xT) regimes when the observations are generated from Tweedie
# distributions.  The input is provided in a list that ordinarily would be geneated by a
# menu-driven or shiny front end.
# INPUT = a list (inputlist) of:
#        xC = N x mC array of observations under a 'control' regime.  Each row of xC is
#             a set of mC realizations from Tweedie distributions whose parameters are
#             realizations from a common population distribution (the parameters for the
#             distributions of the observations in each row of xC need not be the same
#        xT = N x mT array of 'test' observations.  Each row of xT is a set of mT
#             realizations from Tweedie distributions whose parameters are realizations
#             from a common population distribution (like xC), where the population
#             distribution for some of the rows will be the same as for the 'control'
#             regime and some will be from an alternative population different from the
#             'control' regime.  The purpose of the calculations is to identify which
#             rows of xC and xT are from the same population and which are not.
#     inits = starting values for maximum likelihood estimation of prior parameters for
#             the 'control' observation distributions (Empirical Bayes estimation)
#      Tmod = parameters specifying how the alternative population distributions are to
#             be obtained from the 'control' distribution population parameters.
```

```
# TargDiffs = target values at which to calculate the posterior cdf of differences
#             between the 'control' and 'test' observations
#      pi0x = values at which to calculate the density and cdf of the posterior
#             probability that the population parameters are the same for the 'control'
#             and 'test' observations
# ngridpts, prune, digits, print.notes = control parameters
#      misc = a list of assorted quantities defining the run
#
# The calculations use xC and xT as labelled in inputlist also after interchanging them
#
# OUTPUT = collection of lists:
#          inputlist = input to the function
#
#  PostMargDen_out = output from PostMargDen.fn, consisting of PostMargDen_CT and
#                    PostMargDen_TC, each a list consisting of
#            CtlPrms = list of
#                          par = maximum likelihood estimate of the mean of the prior
#                                distn of the transformed Tweedie parameters
#                       covmat = asymptotic covariance matrix of Tweedie parameter
#                                prior distn
#                      corrmat = correlation matrix corresponding to covmat
#                    xi_mu_phi = natural Tweedie parameter values corresponding to
#                                par vals
#              Sumry = list of
#                  rowMargDen = N x 1 matrix of product of joint marginal densities
#                                for each row of xC
#                 rowPostMean = N x 3 matrix of values of the (transformed) posterior
#                                means corresponding to the rows of xC
#                   xi_mu_phi = N x 3 matrix of value of the natural Tweedie
#                                parameters corresponding to rowPostMean
#        MargSum_Co = list of
#                      PostDen = marginal densities
#                     PostMean = posterior expected values, both corresponding to each row of the
#                                xT array when the a priori expected parameter values for each row
#                               of xT are the values in the rowPostMean component of the Sumry list
#        MargSum_Ct = list of
#                      PostDen = marginal densities
#                     PostMean = posterior expected values, both corresponding to each row of the
#                                xT array when the a priori expected parameter values for each row
#                                of xT are the Tmod-driven transforms of the values in the
#                                rowPostMean component of the Sumry list
#       Marg_LikRat = values of the ratios of likelihoods:
#                           (MargSum_Ct$rowMargDen/MargSum_Co$rowMargDen)
#         TransPars = Posterior means of Tweedie parameters for control regime obsns
#
#    ScrnCalcs_out = output from ScrnCalcs.fn = collection of lists:
#       ScrnCalcs_CT = list consisting of
#        Post_Den_CDF_pi0 = list consisting of
#                          Den = posterior density of pi0 = P(same process generated
#                                params for xT & xC) evaluated at arguments in pi0x
#                          CDF = posterior CDF of pi0 evaluated at arguments in pi0x
#                        E_pi0 = posterior expected value of pi0
#                        Ratio = (p-pi0)/pi0 evaluated at arguments in pi0x
#                    DenPlot = graphical display of density generated using ggplot2
#                    CDFPlot = graphical display of cdf generated using ggplot2
#              Pgam0 = list of
#                 Lik_Rat = likelihood ratio (same as MargLikRat)
#              P.gam.eq.0 = P(same process generated xC & xT parameters) for each
#                           xT row
#       ScrnCalcs_TC = list consisting of the same components as ScrnCalcs_CT
#                      except that the roles of xC and xT are reversed
#      Comb_Pgam0 = Combination of the Pgam0 components of ScrnCalcs_CT and
#                   ScrnCalcs_TC as a data frame rather than separate lists
#  Comb_pi0_Plots = list consisting of
#            DenPlot = Posterior densities of pi0 when xC and xT are as
#                      input and after interchanging them
#            CDFPlot = Posterior cdfs of pi0 when xC and xT are as
#                      input and after interchanging them
#      Diffs11_0, Diffs11_1, Diffs11_2, Diffs12_0, Diffs12_1, Diffs12_2
#              = data frames of 1-posterior cdf of differences between the gene expressions
#                when xT and xC are generated by the same process and when they are not: first
```

```
#                  when xC and xT are as input and then when they are interchanged.  In
#                  particular, let X1a, X1b ~ Tweedie with parameters given by xi_mu_phi
#                  from the Sumry list, X2 ~ Tweedie with parameters transformed according to
#                  Tmod.  Calculate 1-cdf of D11 = X1a - X1b and D12 = X2 - X1a at the values in
#                  TargDiffs in 3ways for each row of xT: Diff_Srv_0 = assuming X1 = 0,
#                  Diff_Srv_1 = assuming X1 > 0, and Diff_Srv_2 = averaged over X1
#
#    ScreenEngine_Elapsed.Time = total time for the screen engine calculations
#    PostMargDen_Elapsed.Time = time to calculate the components of PostMargDen_out
#    ScrnCalcs_Elapsed.Time = time to calculate the components of ScrnCalcs_out
#
#                                                     A.L.Gould  December 2023, April 2024
#########################################################################################
#                        TweedieScreenEngine.fn  code starts here                      #
#########################################################################################

  library(doFuture)                        # Set up for parallel processing
  future::plan(multisession,workers=parallel::detectCores()-2)
  require(ggplot2)
  t3 <- t2 <- proc.time()[3]
  PostMargDen_out <- PostMargDen.fn(inputlist)
  PMDT <- Elapsed.time.fn(t3)
  t4 <- proc.time()[3]
  ScrnCalcs_out <- ScrnCalcs.fn(PostMargDen_out, inputlist)
  return(list(Call=sys.call(), Date=date(), ScreenEngine_Elapsed.Time=Elapsed.time.fn(t2),
              PostMargDen_Elapsed.Time=PMDT, ScrnCalcs_Elapsed.Time=Elapsed.time.fn(t4),
              inputlist=inputlist, PostMargDen_out=PostMargDen_out,
              ScrnCalcs_out=ScrnCalcs_out))
}

PostMargDen.fn <- function(inputlist)
{
# Calculates quantities needed for Bayesian analysis of genomic screening assuming that
# observations are generated by Tweedie distributions.  The prior distribution of the
# Tweedie density parameters for the control group are determined by maximum likelihood
# so that the calculationsthat follow are empirical rather than Bayes tp avoid
# specifying the prior distribution parameters directly.  For any set of (xiC,muC,phiC)
# control group parameter values, the corresponding test group parameter values are
# (xiT,muT,phiT) where muT = Tmod[2]+muC, phiT=Tmod[3]*phiC, and xiT is determined from
# Tmod[1] = (xiT-1)*(2-xiC)/(2-xiT)*(xiC-1).  The apparent redundancy of the output
# allows the roles of the observation matrices from the 'control' and 'test' regimes to
# be interchanged so that evaluation of both overexpression and underexpression of
# genes can be accommodated by the same analytic approach.
#
# INPUT  -- a list (inputlist) containing:
#              xC = matrix of control group observations
#              xT = matrix of test group observations
#           inits = initial values of natural Tweedie distribution parameters xi and
#                   phi (initial value of mu is mean of positive values in xC)
#            Tmod = change specifications for test gp parameters
#        ngridpts = no. of grid points along each dimension of params for quadrature
#           prune = fraction of extreme quadrature points that can
#                   be removed without materially affecting precision
# OUTPUT
#   PostMargDen_CT = PostMargDenstep.fn output with original definitions of 'control'
#                    and 'test'
#   PostMargDen_TC = PostMargDenstep.fn output with 'control' and 'test' interchanged

  tsim2.fn <- function(x, inputlist)
  {
    if (x==1) y <- PostMargDenstep.fn(inputlist$xC, inputlist$xT, inputlist)
    if (x==2) y <- PostMargDenstep.fn(inputlist$xT, inputlist$xC, inputlist)
    return(y)
  }

  y <- foreach::foreach(x=1:2,.options.future = list(seed = TRUE)) %dofuture% tsim2.fn(x,
                                                                               inputlist)
  return(list(call=sys.call(), date=date(), PostMargDen_CT = y[[1]], PostMargDen_TC = y[[2]]))
}

PostMargDenstep.fn <- function(xC, xT, inputlist)
```

```
{
#  OUTPUT
#     CtlPrms = prior distribution parameters for transformed Tweedie distn
#               parameters from t_Tweedie_MLCalcs.fn assuming that the same
#               Tweedie distn applies for each element of the xC obsn matrix
#       Sumry = list containing marginal likelihoods (rowMargDen) and corresonding posterior
#               expected values (rowPostMean) of transformed Tweedie density parameters for
#               each row of the xC matrix when the prior distribution of the density parameters
#               is multivariate normal with mean and covariance matrix determined
#               by the maximum likelihood estimates from t_Tweedie_MLCalcs.fn
#  MargSum_Co = list containing marginal densities (rowMargDen) and posterior expected parameter
#               values (rowPostMean) corresponding to each row of the xT matrix when the a
#               priori expected parameter values for each row are the values provided in the
#               rowPostMean component of the Sumry_C list
#  MargSum_Ct = list containing marginal densities (rowMargDen) and posterior expected parameter
#               values (rowPostMean) corresponding to each row of the xT matrix when the a
#               priori expected parameter values for each row are the transformed values of the
#               rowPostMean component of the Sumry_C list applying the Tmod argument values
# Marg_LikRat = values of the ratios of likelihoods:
#                 Marg_LikRat = MargSum_Ct$rowMargDen/MargSum_Co$rowMargDen
#  TransPars = Posterior means of Tweedie parameters for control regime obsns
#
#                                                      A.L.Gould  October 2023

 tsim1.fn <- function(x, xT, Sumry, CtlPrms, TransPars, prune)
 {
   if (x==1)
   {
     arg1 <- Sumry$rowPostMean
     arg2 <- CtlPrms$covmat
   }
   else
   {
     arg1 <- TransPars$new_pars
     arg2 <- TransPars$newcov
   }
   return(Tweedie_PostDenMean.fn(xT, arg1, arg2, prune))
 }

 t0 <- proc.time()[3]
 ##########################################################################
 #    Step 1: get ML estimates of model parameters for ctl gp and for test group   #
 ##########################################################################
 CtlPrms <- Tweedie_MLCalcs.fn(xC, inits=inputlist$inits)

 ##########################################################################
 #    Step 2: Get marginal densities and posterior means of transformed parameters #
 #            based on 'control' regime observations                              #
 ##########################################################################

 Sumry <- Tweedie_Sumry.fn(CtlPrms, xC, prune=inputlist$prune)

 TransPars <- Tweedie_trans.fn(Sumry, inputlist$Tmod)

 ##########################################################################
 #     Step 3: Using the findings from Step 2 as prior information, get marginal   #
 #             likelihoods and posterior expected parameter values for each test   #
 #             observation matrix row assuming that the same parameters apply as   #
 #             for the corresponding control observation matrix.                   #
 ##########################################################################
 ##########################################################################
 #     Step 4: Alter the posterior expected parameter values from Step 2 as        #
 #             specified by Tmod.  Repeat Step 3 with the changed parameter values #
 ##########################################################################

 y <- foreach::foreach(x=1:2,.options.future = list(seed = TRUE)) %dofuture% tsim1.fn(x,
                                    inputlist$xT,Sumry,CtlPrms,TransPars,inputlist$prune)
 MargSum_Co <- y[[1]]                # Step 3
 MargSum_Ct <- y[[2]]                # Step 4
 Marg_LikRat <- MargSum_Ct$Post_Den/MargSum_Co$Post_Den
 if(inputlist$print.notes==TRUE) cat("\nPostMargDenStep: ", Elapsed.time.fn(t0))
```

```
  return(list(CtlPrms=CtlPrms, Sumry=Sumry, MargSum_Co=MargSum_Co,
              MargSum_Ct=MargSum_Ct, TransPars=TransPars, Marg_LikRat=Marg_LikRat))
}

Tweedie_MLCalcs.fn <- function(X, inits=NULL)
{
# Calculates maximum likelihood estimates of transformed Tweedie distn parameters
# INPUT
#      inits = initial values of natural Tweedie parameters xi, and phi
#              (mu is calculated as overall mean of positive X values)
#          X = set of observations
# OUTPUT
#        par = expected mean of the maximum likelihood estimates of
#              transformed tweedie parameters
#     covmat = asymptotic covariance matrix for the parameters
#              corresponding to the values in par
#    corrmat = correlation matrix corresponding to covmat
#  xi_mu_phi = natural tweedie parameter values corresponding to values in par

  tot_LL.fn <- function(pars,x) return(-sum(log(t_Tweedie_den.fn(pars,x))))

  t1 <- proc.time()[3]
  par0 <- c(gtools::logit(inits[1]-1),log(mean(X[X>0])),log(inits[2]))
  f <- optim(par0, tot_LL.fn, x=as.vector(t(X)), hessian=T)
  xi_mu_phi <- c(gtools::inv.logit(f$par[1],1,2), exp(f$par[2]), exp(f$par[3]))
  covmat <- solve(f$hessian)
  return(list(call=sys.call(), Date=date(), Elapsed.Time=Elapsed.time.fn(t1),
              pars=f$par, xi_mu_phi=xi_mu_phi, covmat=covmat,
 corrmat=cov2cor(covmat)))
}

Tweedie_Sumry.fn <- function(CtlPrms, xC, prune=NULL)
{
# Calculates marginal densities of control regime observations and
# posterior means of (transformed) tweedie distribution parameters
# INPUT
#   CtlPrms = output list from Tweedie_MLCalcs.fn
#        XC = N x M matrix of control regime observations
# OUTPUT
#  rowMargDen = N x 1 matrix of product of joint marginal densities for each row of xC
# rowPostMean = N x 3 matrix of values of the (transformed) posterior
#               means corresponding to rows of the control regime obsns
#                                                          A.L.Gould  October 2023
  fa.fn <- function(par, x)
  {
    y <- t_Tweedie_den.fn(par, as.vector(t(x)))
    dim(y) <- rev(dim(x))
    return(apply(t(y),1,prod))
  }

  fb.fn <- function(denvec,pts,wts) return((as.vector(denvec)*wts)%*%pts)

  t1 <- proc.time()[3]
  qgrd <- mgauss.hermite.fn(10, CtlPrms$pars, CtlPrms$covmat, prune=prune)
  y <- apply(vec2mat.fn(qgrd$points), 1, fa.fn, x=vec2mat.fn(xC,2))
  MargDen <- y %*% qgrd$weights
  u <- lapply(asplit(y,1), fb.fn, pts=qgrd$points, wts=qgrd$weights)
  uu <- unlist(u)
  dim(uu) <- c(3,length(u))
  uu <- t(uu)/as.vector(MargDen)
  xi_mu_phi <- cbind(gtools::inv.logit(uu[,1])+1, exp(uu[,2:3]))
  return(list(call=sys.call(), Date=date(), Elapsed.Time=Elapsed.time.fn(t1),
              rowMargDen=MargDen, rowPostMean=uu, xi_mu_phi=xi_mu_phi))
}

Tweedie_trans.fn <- function(Sumry, Tmod=NULL)
{
#  Original transformed Tweedie distribution parameters are columns of pars. This function
#  returns transformed parameters for the test regime by modifying the natural Tweedie
#  distribution parameters for the control regime according to the elements of Tmod.
#  Tmod[1] amounts to the (new) odds ratio for the xi parameter after subtracting 1
```

```
#  ( 1 < xi < 2). Tmod[2] is a shift in the mean parameter mu.  Tmod[3] is a multiple of the
#  precision parameter phi.  Natural parameters (xi, mu, phi) corresponding to pars are obtained
#  as  xi = 1 + 1/(1 + exp(-pars[,1])), mu = exp(pars[,2]), phi = exp(pars[,3])
#  The modified natural parameter values xiT, muT, phiT are obtained as
#    xiT = ((2*Tmod[1]-1)*xi + 2*(1-Tmod[1]))/((Tmod[1]-1)*xi + 2-Tmod[1]),
#    muT = mu + Tmod[2], phiT = Tmod[3]*phi. The transformed modified natural
#  parameters are obtained as new_pars[,1] <- logit(xiT - 1),
#  new_pars[2] = log(muT), new_pars[3] = log(phiT).
#
#  INPUT     Sumry = output list from Tweedie_Sumry.fn
#  OUTPUT list containing:
#        new_pars = new transformed natural parameters after applying Tmod specifications
#   new_xi_mu_phi = new natural parameters
#          newcov = empirical covariance matrix of the new (transformed) parameters
#                                                              A. L. Gould  October 2023
  pars <- vec2mat.fn(Sumry$rowPostMean)
  xi_mu_phi <- cbind(1+gtools::inv.logit(pars[,1]), exp(pars[,2:3]))
  new_xi <- ((2*Tmod[1]-1)*xi_mu_phi[,1] + 2*(1-Tmod[1]))/
                          ((Tmod[1]-1)*xi_mu_phi[,1] + 2-Tmod[1])
  new_xi_mu_phi <- cbind(new_xi, cbind(xi_mu_phi[,2]+Tmod[2], Tmod[3]*xi_mu_phi[,3]))
  newpars <- cbind(gtools::logit(new_xi_mu_phi[,1]-1), log(new_xi_mu_phi[,2:3]))
  newcov <- cov(newpars)
  return(list(new_pars=newpars, new_xi_mu_phi=new_xi_mu_phi, newcov=newcov))
}

t_Tweedie_den.fn <- function(par,x)
     return(tweedie::dtweedie(x,gtools::inv.logit(par[1],1,2),exp(par[2]),
                             exp(par[3])))

mgauss.hermite.fn <- function(n, mu, sigma, prune=NULL)
{
## compute multivariate Gaussian quadrature points
## n     - number of points each dimension before pruning
## mu    - mean vector
## sigma - covariance matrix
## prune - NULL - no pruning; [0-1] - fraction to prune

  if(!all(dim(sigma) == length(mu)))
      stop("mu and sigma have nonconformable dimensions")
  dm  <- length(mu)
  gh  <- ecoreg::gauss.hermite(n)
                          #  idx grows exponentially in n and dm
  idx <- as.matrix(expand.grid(rep(list(1:n),dm)))
  pts <- matrix(gh[idx,1],nrow(idx),dm)
  wts <- apply(matrix(gh[idx,2],nrow(idx),dm), 1, prod)
  if(!is.null(prune))                # prune
  {
    qwt <- quantile(wts, probs=prune)
    pts <- pts[wts > qwt,]
    wts <- wts[wts > qwt]
  }
            # rotate, scale, translate points
  eig <- eigen(sigma)
  rot <- eig$vectors %*% diag(sqrt(eig$values))
  pts <- t(rot %*% t(pts) + mu)
  return(list(points=pts, weights=wts))
}

ScrnCalcs.fn <- function(PostMargDen_out, inputlist)
{
  tsim3.fn <- function(x, PostMargDen, rnxc)
  {
    if (x==1) return(ScrnCalcStep.fn(PostMargDen$PostMargDen_CT, inputlist, rnxc))
    if (x==2) return(ScrnCalcStep.fn(PostMargDen$PostMargDen_TC, inputlist, rnxc))
  }

  rnxc <- rownames(inputlist$xC)
  pi0x <- eval(parse(text=inputlist$pi0x))
#                                                      parallel processing
  y <- foreach::foreach(x=1:2,.options.future = list(seed = TRUE))  %dofuture% tsim3.fn(x,
                                                                PostMargDen_out,rnxc)
```

```
  Diffs11_0 <- cbind(y[[1]]$Diffs_Srv11$Diff_Srv_0, y[[2]]$Diffs_Srv11$Diff_Srv_0)
  Diffs12_0 <- cbind(y[[1]]$Diffs_Srv12$Diff_Srv_0, y[[2]]$Diffs_Srv12$Diff_Srv_0)
  Diffs11_1 <- cbind(y[[1]]$Diffs_Srv11$Diff_Srv_1, y[[2]]$Diffs_Srv11$Diff_Srv_1)
  Diffs12_1 <- cbind(y[[1]]$Diffs_Srv12$Diff_Srv_1, y[[2]]$Diffs_Srv12$Diff_Srv_1)
  Diffs11_2 <- cbind(y[[1]]$Diffs_Srv11$Diff_Srv_2, y[[2]]$Diffs_Srv11$Diff_Srv_2)
  Diffs12_2 <- cbind(y[[1]]$Diffs_Srv12$Diff_Srv_2, y[[2]]$Diffs_Srv12$Diff_Srv_2)
  Comb_Pgam0 <- cbind(y[[1]]$Pgam0, y[[2]]$Pgam0)
  names(Comb_Pgam0) <- c("Lik_Rat_CT","P.gam.eq.0_CT","Lik_Rat_TC","P.gam.eq.0_TC")
  Comb_pi0_Plots <- Comb_pi0_Plots.fn(pi0x, y[[1]]$Post_Den_CDF_pi0, y[[2]]$Post_Den_CDF_pi0)
  return(list(call=sys.call(), date=date(), ScrnCalcs_CT=y[[1]], ScrnCalcs_TC=y[[2]],
              Comb_Pgam0 = Comb_Pgam0, Comb_pi0_Plots=Comb_pi0_Plots,
              Diffs11_0=Diffs11_0, Diffs11_1=Diffs11_1, Diffs11_2=Diffs11_2,
              Diffs12_0=Diffs12_0, Diffs12_1=Diffs12_1, Diffs12_2=Diffs12_2))
}

ScrnCalcStep.fn
function(PostMarg, inputlist, rnxc)
{
  tsim4.fn <- function(x, inputlist, xi_mu_phi1, xi_mu_phi2, rnxc)
  {
    if (x==1) y <- Tweedie_Diffs_Srv.fn(cbind(xi_mu_phi1, xi_mu_phi1), inputlist$TargDiffs,
                                      rnxc, digits=inputlist$digits)
    if (x==2) y <- Tweedie_Diffs_Srv.fn(cbind(xi_mu_phi1, xi_mu_phi2), inputlist$TargDiffs,
                                      rnxc, digits=inputlist$digits)
    return(y)
  }

  t1 <- proc.time()[3]
  pi0x <- eval(parse(text=inputlist$pi0x))
  xi_mu_phi1 <- PostMarg$Sumry$xi_mu_phi
  xi_mu_phi2 <- PostMarg$TransPars$new_xi_mu_phi
  y <- foreach::foreach(x=1:2,.options.future = list(seed = TRUE)) %dofuture% tsim4.fn(x,
                                                   inputlist, xi_mu_phi1, xi_mu_phi2, rnxc)
  Post_Den_CDF_pi0 <- Post_Den_CDF_pi0.fn(pi0x, PostMarg$Marg_LikRat)
  DenCDF_pi0 <- as.data.frame(cbind(pi0x, Post_Den_CDF_pi0$Den, Post_Den_CDF_pi0$CDF))
  names(DenCDF_pi0) <- c("pi0", "f_pi0","F_pi0")
  DenPlot <- ggplot(data=DenCDF_pi0,aes(x=pi0)) + theme_bw() +
               geom_line(aes(y=f_pi0), linewidth=1.1) +
               xlab(expression(pi [0])) + ylab(expression(f( pi [0] ))) +
               theme(axis.title.x=element_text(size=14)) +
               theme(axis.title.y=element_text(angle=0,size=14,vjust=.5)) +
               theme(axis.text.x=element_text(size=11)) +
               theme(axis.text.y=element_text(size=11))
  CDFPlot <- ggplot(data=DenCDF_pi0, aes(x=pi0)) + theme_bw() +
               geom_line(aes(y=F_pi0), linewidth=1.1) +
               xlab(expression(pi [0])) + ylab(expression(F( pi [0] ))) +
               theme(axis.title.x=element_text(size=14)) +
               theme(axis.title.y=element_text(angle=0,size=14,vjust=.5)) +
               theme(axis.text.x=element_text(size=11)) +
               theme(axis.text.y=element_text(size=11))

  Pgam0 <- Screen_Probs.fn(inputlist, PostMarg$Marg_LikRat, Post_Den_CDF_pi0, rnxc)

  if(inputlist$print.notes==TRUE) cat("\nScrnCalcStep: ", Elapsed.time.fn(t1))
  return(list(call=sys.call(), date=date(), Post_Den_CDF_pi0=Post_Den_CDF_pi0,
              DenPlot=DenPlot, CDFPlot=CDFPlot, Pgam0=Pgam0, Diffs_Srv11=y[[1]],
              Diffs_Srv12=y[[2]]))
}

Post_Den_CDF_pi0.fn <- function(pi0, MargLiks, zeta=5)
{
  f0.fn <- function(pi0, R, zeta)
    return(exp((length(R) + zeta - 1)*log(pi0) + sum(log(1 + ((1-pi0)/pi0)*R))))

  Den <- sapply(pi0, f0.fn, R=MargLiks, zeta=zeta)
  H <- sum(Den)*(pi0[2]-pi0[1])
  Den <- Den/H
  CDF <- cumsum(Den)/sum(Den)
  return(list(Den=Den, CDF=CDF, E_pi0=1-(pi0[2] - pi0[1])*sum(CDF), Ratio=(1-pi0)/pi0))
}
```

```
Tweedie_Diffs_Srv.fn <- function(ximuphi12, diffs, rnxc, eps=1e-8, digits=NULL)
{
# Let X1 ~ Tweedie(ximuphi1) and X2 ~ Tweedie(ximuphi2), where ximuphi1 and ximuphi2
# could be the same or different 3-elt vectors.  This function calculates the values
# of the cdf of the difference D = X2 - X1 overall and assuming X1 = 0 or X1 > 0.
# INPUT
#
# ximuphi12 = cbind(ximuphi1, ximuphi2) = N x 6 matrix of 'natural' Tweedie
#             distribution parameters for X1 & X2
#     diffs = vector of positive difference values
#      rmxc = row names of xC (names of genes)
#
# OUTPUT = list of 3 N x 3*length(diffs) matrices whose rows are the probabilities that
#          the true difference values exceed the values in diffs:
#  Diff_Srv_0 = matrix assuming X1 = 0
#  Diff_Srv_1 = matrix assuming X1 > 0
#  Diff_Srv_2 = matrix averaged over X1
#                                                               A.L.Gould  October 2023
  intgrnd.fn <- function(x, dif, ximuphi12)
  {
    x1 <- tweedie::ptweedie(x+dif, xi=ximuphi12[4], mu=ximuphi12[5], phi=ximuphi12[6])
    x2 <- tweedie::dtweedie(x, xi=ximuphi12[1], mu=ximuphi12[2], phi=ximuphi12[3])
    return(x1*x2)
  }

  diff0_cdf.fn <- function(ximuphi12, diffs)
    return(as.vector(tweedie::ptweedie(diffs, xi=ximuphi12[4], mu=ximuphi12[5],
                                   phi=ximuphi12[6])))

  Srv0.fn <- function(ximuphi12, diffs)
    return(1-apply(ximuphi12, 1, diff0_cdf.fn, diffs=diffs))

  cdf1.fn <- function(diff, ximuphi12)
    return(integrate(intgrnd.fn, 0, Inf, dif=diff, ximuphi12=ximuphi12)$value)

  diff1_cdf.fn <- function(diffs, ximuphi12)
    return(sapply(diffs, cdf1.fn, ximuphi12=ximuphi12))

  Srv1.fn <- function(ximuphi12, diffs)
      return(1-apply(ximuphi12, 1, diff1_cdf.fn, diffs=diffs))

  cdf2.fn <- function(diff, ximuphi12)
  {
    p0 <- tweedie::dtweedie(0, xi=ximuphi12[1], mu=ximuphi12[2],
                            phi=ximuphi12[3])
    yy <- p0*tweedie::ptweedie(diff, xi=ximuphi12[4], mu=ximuphi12[5],
                               phi=ximuphi12[6])
    return(yy + integrate(intgrnd.fn, 0, Inf, dif=diff, ximuphi12=ximuphi12)$value)
  }

  diff2_cdf.fn <- function(diffs, ximuphi12)
                 return(sapply(diffs, cdf2.fn, ximuphi12=ximuphi12))

  Srv2.fn <- function(ximuphi12, diffs)
                 return(1-apply(ximuphi12, 1, diff2_cdf.fn, diffs=diffs))

#####################################################################################
#                        Tweedie_Diffs_Srv.fn code starts here                      #
#####################################################################################

t5 <- proc.time()[3]
Diff_Srv_0 <- round(as.data.frame(t(Srv0.fn(ximuphi12, diffs))), digits)
Diff_Srv_1 <- round(as.data.frame(t(Srv1.fn(ximuphi12, diffs))), digits)
Diff_Srv_2 <- round(as.data.frame(t(Srv2.fn(ximuphi12, diffs))), digits)
names(Diff_Srv_0) <- names(Diff_Srv_1) <- names(Diff_Srv_2) <- paste0("d=",diffs)
rownames(Diff_Srv_0) <- rownames(Diff_Srv_1) <- rownames(Diff_Srv_2) <- rnxc
if(inputlist$print.notes==TRUE) cat("\nSDiffsSrv: ", Elapsed.time.fn(t5))
return(list(call=sys.call(), Date=date(), Diff_Srv_0=Diff_Srv_0,
            Diff_Srv_1=Diff_Srv_1, Diff_Srv_2=Diff_Srv_2))
}
```

```
Comb_pi0_Plots.fn <- function(pi0x, PostDist_pi0_CT, PostDist_pi0_TC)
{
# Plot the posterior density of pi0 for CT and TC on the same plot,
# plot the posterior cdf of pi0 for CT and TC on the same plot
#                                                      A.L.Gould December 2023
  xx <- as.data.frame(cbind(pi0x, PostDist_pi0_CT$Den, PostDist_pi0_TC$Den,
                            PostDist_pi0_CT$CDF, PostDist_pi0_TC$CDF))
  names(xx) <- c("pi0x","DenCT","DenTC","CDFCT","CDFTC")
  DenPlot_Comb <- ggplot(data=xx) + theme_bw() +
                    theme(panel.grid.major=element_blank(), panel.grid.minor=element_blank()) +
                    geom_line(aes(x=pi0x, y=DenCT, linetype="DenCT"), linewidth=1.1) +
                    geom_line(aes(x=pi0x,y=DenTC, linetype="DenTC"), linewidth=1.1) +
                    scale_linetype_manual(values = c("solid", "dashed"),
                      guide = guide_legend(title = "")) + labs(linetype = "") +
                    xlab(expression(pi [0])) + ylab(expression(f( pi [0] ))) +
                    theme(legend.text=element_text(size=10),legend.position.inside=c(.18,.8),
                          legend.key.width=unit(1,"cm")) +
                    theme(axis.title.x=element_text(size=14)) +
                    theme(axis.title.y=element_text(angle=0,size=14,vjust=.5)) +
                    theme(axis.text.x=element_text(size=11)) +
                    theme(axis.text.y=element_text(size=11))
  CDFPlot_Comb <- ggplot(data=xx) + theme_bw() +
                    theme(panel.grid.major=element_blank(), panel.grid.minor=element_blank()) +
                    geom_line(aes(x=pi0x, y=CDFCT, linetype="CDFCT"), linewidth=1.1) +
                    geom_line(aes(x=pi0x, y=CDFTC, linetype="CDFTC"), linewidth=1.1) +
                    scale_linetype_manual(values = c("solid", "dashed"),
                      guide = guide_legend(title = "")) + labs(linetype = "") +
                    xlab(expression(pi [0])) + ylab(expression(F( pi [0] ))) +
                    theme(legend.text=element_text(size=10),legend.position.inside=c(.18,.8),
                          legend.key.width=unit(1,"cm")) +
                    theme(axis.title.x=element_text(size=14)) +
                    theme(axis.title.y=element_text(angle=0,size=14,vjust=.5)) +
                    theme(axis.text.x=element_text(size=11)) +
                    theme(axis.text.y=element_text(size=11))
  return(list(DenPlot_comb=DenPlot_Comb, CDFPlot_comb=CDFPlot_Comb))
}

Elapsed.time.fn <- function (t1)
{
    tt <- proc.time()[3] - t1
    H <- floor(tt/3600)
    M <- floor((tt - H * 3600)/60)
    ifelse (H > 1, x1 <- " hours", x1 <- " hour")
    ifelse (M > 1, x2 <- " minutes", x2 <- " minute")
    S <- round(tt - 3600 * H - 60 * M, 2)
    w <- paste(S, "seconds")
    if ((H > 0) && (M > 0))  w <- paste(H, x1, ", ", M, x2, ", and ", w, sep = "")
    if ((H > 0) && (M < 1))  w <- paste(H, x1, " and ", w, sep = "")
    if ((H < 1) && (M > 0))  w <- paste(M, x2, " and ", w, sep = "")
    w <- paste("Elapsed time:", w)
    return(w)
}

vec2mat.fn <- function(v,opt=1)
{
    #  If v is a vector with m elements, then converts v to a 1 x m array (opt=1)
    #  or an m x 1 array (opt=2). If v is a matrix, then v is returned unchanged.
    if (length(dim(v)) > 1) v <- matrix(as.matrix(v),nrow=dim(v)[1],ncol=dim(v)[2])
    else
    {
        v <- matrix(v,nrow=1,ncol=length(v))
        if (opt > 1) v <- t(v)
    }
    return(v)
}
```